\documentclass[letterpaper, 10 pt, conference]{ieeeconf}  
\usepackage{graphicx}
\usepackage{multirow}

\IEEEoverridecommandlockouts                              

\title{\LARGE \bf
Rapid Modeling Architecture for Lightweight Simulator to Accelerate and Improve Decision Making for Industrial Systems
}

\author{Takumi Kato and Zhi Li Hu%
\thanks{Takumi Kato (Correspondence: {\tt\small takumi.kato@ieee.org}) and Zhi Li Hu ({\tt\small Zhi.Hu@hal.hitachi.com}) are with the R\&D Group, Hitachi America, Ltd., Santa Clara, California, 95054, USA. }%
}

\begin{document}

\maketitle
\thispagestyle{empty}
\pagestyle{empty}

\begin{abstract}
Designing industrial systems, such as building, improving, and automating distribution centers and manufacturing plants, involves critical decision-making with limited information in the early phases. The lack of information leads to less accurate designs of the systems, which are often difficult to resolve later. It is effective to use simulators to model the designed system and find out the issues early. 
However, the modeling time required by conventional simulators is too long to allow for rapid model creation to meet decision-making demands.
In this paper, we propose a Rapid Modeling Architecture (RMA) for a lightweight industrial simulator that mitigates the modeling burden while maintaining the essential details in order to accelerate and improve decision-making. We have prototyped a simulator based on the RMA and applied it to the actual factory layout design problem. We also compared the modeling time of our simulator to that of an existing simulator, and as a result, our simulator achieved a 78.3\% reduction in modeling time compared to conventional simulators.
\end{abstract}

\section{INTRODUCTION}
Designing industrial systems, such as distribution centers and manufacturing plants classified as discrete manufacturing and equivalent systems, requires critical decision-making under conditions of uncertainty or limited information. Early phases often lack the complete data needed to accurately estimate throughput, layout feasibility, and other key factors. These stages involve layout design, process development, machinery selection, setting performance targets, estimating costs, and more. Errors made during these initial steps can be costly to resolve later, as the scale and complexity of industrial operations leave little room for post-implementation changes. Design projects may involve building new systems from scratch or improving existing ones through System Integration (SI) and automation.

Simulation technology is valuable for verifying design assumptions and identifying bottlenecks before finalizing costly decisions. By creating simulation models of systems, decision-makers can experiment with different configurations and scenarios to evaluate the impact on operations. Conventional industrial simulators, however, typically require considerable time and specialized modeling knowledge to prepare such models. This modeling time can slow the design iteration process, especially in common scenarios where decision-makers must make prompt decisions to finalize the purchase orders to the partnering vendors.

This paper proposes a Rapid Modeling Architecture (RMA) for lightweight industrial simulators to accelerate and improve decision-making to achieve better and more trustworthy designs of industrial systems. We aim to reduce the modeling burden and accelerate the design-and-analysis feedback loop without sacrificing the overall accuracy and reliability of the simulation. This balance is achieved through the unique abstracting of the industrial system's components, detailed yet easy-to-prepare task descriptions, and the built-in task processing mechanism.

We have implemented a prototype simulator based on the proposed RMA architecture and applied it to a factory layout design problem. Our results show that our approach effectively reduces the modeling time required to build the simulation model. Compared to a widely used conventional industrial simulator, our simulator achieved a 78.3\% reduction in modeling time, indicating significant acceleration and improvement in early-phase decision-making.

The remainder of this paper is organized as follows. Section II reviews the relevant literature and conventional simulators and discusses the main limitations of current industrial simulation tools. Section III details the design of our rapid modeling architecture and highlights its key features. Section IV presents our prototype implementation. Section V explains the modeling experiment and the evaluation results. Finally, Section VI concludes the paper and outlines directions for future work.

\section{RELATED WORK}
We have focused on discrete-event systems and their simulations \cite{law2024simulation}, as discrete manufacturing and equivalent systems are our primary focus.

\subsection{Proprietary Simulators}
We have tested and surveyed the existing simulators that could be used for quick and deep analyses of industrial systems in the early phases. Some of the well-known proprietary industrial simulators are Process Simulate \cite{guerrero2014virtual}, Plant Simulation \cite{siderska2016application}, AnyLogic \cite{muravev2021multiagent}, and Visual Components \cite{visualcomponents2025}. Process Simulate and Plant Simulation excel at highly detailed simulations, but they had a steep learning curve and substantial modeling time for the early phases. AnyLogic and Visual Components were more lightweight options, and they are often used in early quotation phases for some projects in the industry. However, although they offer solid analysis functions, such as visualizations and Key Performance Indicator (KPI) calculation functions, the modeling span was still not short enough to keep up with the decision-making timeline. The existing proprietary tools are more accurate and aesthetic-oriented than needed in the early phases of industrial system design. 

\subsection{Academic Approaches to Shorten the Modeling Time}
Turning to academic approaches and tools, NetLogo \cite{koid2025agent} is a good candidate as it offers a simplified modeling experience, and it was easy to quickly set up a reasonably complex model. However, the system performance analysis modeling requires more complexity regarding agent behavior modeling, which then requires extra time. There were other approaches to reduce the modeling time \cite{kato2020multiagent, suemitsu2022fast, ferrari2024metamodel, hazard2006alphabet}. \\
\cite{kato2020multiagent} offers a simplified multi-agent system simulation modeling. However, it does not cover the modeling of the environment of the agents. Surrogate models \cite{suemitsu2022fast} and meta-models \cite{ferrari2024metamodel} are proposed to avoid the conventional modeling process, such as coding. However, the accuracy can fluctuate because of the nature of the non-linear function approximation, such as neural networks, which require fine-tuning based on domain expertise. The authors in \cite{hazard2006alphabet} have prepared a general simulation and optimization framework for multi-vehicle systems such as AGV-based fulfillment tasks. This tool can quickly analyze such systems, but most projects require a wider range of analyses. The conventional academic approaches are either too abstract or require special tuning skills for the desired analyses in our focus. 

\subsection{Additional Impacts of Shortening Modeling Time}
Simulation models of industrial systems are necessary to utilize optimization techniques, and shortening the modeling time can contribute to better utilization of such techniques. \cite{chen2023multifidelity} presents a unique framework for incorporating simulation models with multiple fidelity to accelerate the optimization computation. For this approach, users still need to prepare some models to start, and that modeling time should be shortened to accelerate the decision-making. This research \cite{yoshitake2023impact} proposes multi-agent reinforcement learning for optimizing an industrial system's operation, specifically a warehouse operation. To train reinforcement learning agents, users still need to prepare simulation models, and the modeling time should be shortened to gain the trained model faster and optimize the warehouse. There are other simulation-based approaches for manufacturing systems \cite{wang2008simulation, mollaghasemi1994multicriteria, diaz2021optimizing, feng2020simulation, lin2022integrated} as well as food services such as restaurants \cite{huan2011simulation}. Over the course of optimization, \cite{diaz2022enabling} also offers knowledge discovery to deepen the understanding of the target system. Overall, shortening modeling time can accelerate the optimization process by leveraging these approaches, which contributes to accelerating and improving decision-making for industrial systems. Since modeling time is often a bottleneck in the development of reinforcement learning (RL) agents, reducing it can substantially accelerate the overall process of RL-based industrial optimization.

\subsection{Positioning of Our Approach}
In this paper, we aim to strike a good balance between proprietary simulators and the conventional academic approaches to shorten modeling time, not be too accurate and aesthetics-oriented (appropriately compromised), and simplify architecture while maintaining the essential details that substantially contribute to the value of results. The success of our approach can lead to the acceleration of utilizing various and powerful optimization techniques in the existing methods.

\section{RAPID MODELING ARCHITECTURE (RMA) FOR LIGHTWEIGHT INDUSTRIAL SIMULATOR}
\subsection{Target Problem}
Our target problem is reducing the modeling time of industrial simulators while preserving essential details that ensure the value of simulation results. Our approach involves designing a Rapid Modeling Architecture (RMA) that simplifies the modeling process by systematically refining essential simulation components and input data based on insights gained from multiple simulation modeling projects. Since industrial systems consist of multiple interacting entities, we designed a multi-agent simulator architecture with simplified environment modeling capability. We narrowed down the following requirements for the RMA to resolve the target problem.
\begin{itemize}
\item (R1) {\it Reducing unnecessary modeling tasks}: omitting unnecessary aesthetics-related tasks and too high Level of Detail (LoD) on spatial representational capabilities.
\item (R2) {\it High representational capability of system behavior}: while it is essential to omit unnecessary LoD, it is vital to have a fine-grained agent (workers, robots, etc.) behavior representation. For example, it is insufficient to merely represent the flow of materials; it is also necessary to specify the entities responsible for transportation, the conditions under which transportation occurs, and other relevant factors.
\item (R3) {\it Input data structure designed for seamless data integration}: Assessing industrial system designs typically requires operational data and product information. While the format of data can vary among industrial systems, there are common structures across the data that can be utilized in the design assessment. These types of data are commonly used among industrial system designs: item location data, logs or commands of transportation (e.g., customer orders for shipping, orders at restaurants, etc.), and product/outcome structure data such as Bill of Materials (BOM).
\end{itemize}

\subsection{Essence of Rapid Modeling Architecture (RMA)}
We designed the RMA to meet the requirements (R1-3). The essence of our solution can be described as follows:
\begin{itemize}
\item (S1) Voxel-based distilled simulation space architecture to abstract the industrial systems to avoid overly detailed dimensions measurements and component modeling
\item (S2) Streamlined input data structure and processing mechanism to represent and deal with detailed behavior modeling with simplified simulation space architecture 
\end{itemize}
While (S1) simplifies the spatial representation of the industrial systems, (S2) ensures that agent behavior and operational data are accurately captured and processed.

 (S1) reduces unnecessary aesthetic details while preserving essential structural elements needed for accurate simulations, which fulfills (R1). It focuses on the architecture of the simulation space, abstracting industrial systems to streamline modeling. We adopted a voxel-based representation to simplify component placements and a grid-based simplification for efficient computation. We distilled the minimal necessary components to ensure accurate output generation based on our past simulation-related projects in various fields. We narrowed down to three types of components to represent the industrial systems for simplification, namely, agents, receptors, and items. Unlike conventional simulators, users do not need to select from various options to model their systems. Also, since the space is represented as voxels, it is easy to decide where to place objects based on the approximation of dimensions. The details of (S1) are explained in subsection C.

(S2) is the input data structure and processing mechanism that provides the high representational capability of system behavior and seamless data integration, which fulfills (R2) and (R3). We primarily use three types of input data for simulations: (i) layout and basics data, (ii) work orders for agents, and (iii) item location data. There are two types of work orders: (a) transportation work orders and (b) assembly work orders, because tasks in the industrial systems can generally be categorized as either transporting items from point A to point B, or assembling and disassembling items. These two work orders help users detail model behaviors to gain results that contribute to the design assessment. (a) and (b) can also be omitted based on the users' needs. Subsection D details the data structures of (S2), while subsection E explains the processing mechanism. 

\subsection{(S1) Voxel-based Three-Element Simulation Space Architecture}
\begin{figure}[tb]
\centering
\parbox{\textwidth}{
\includegraphics[scale=0.4]{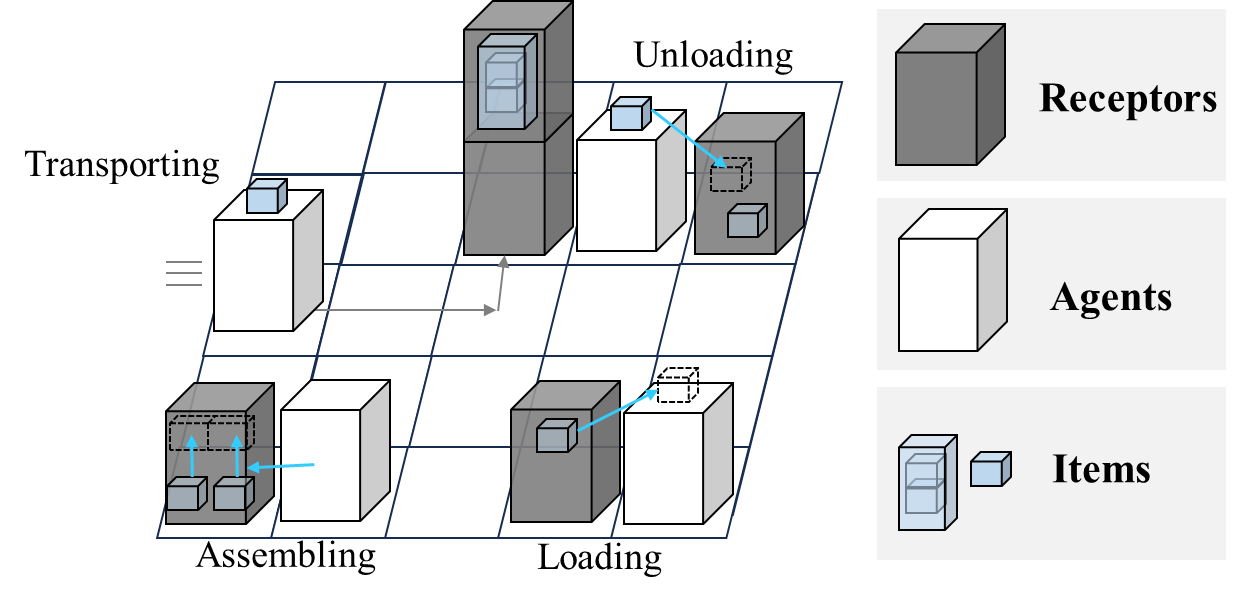}
}
\caption{Voxel-based three-element simulation space. This figure illustrates the three component types: Receptors, Agents, and Items. Items represent processed materials in the simulated industrial system. Receptors store items, while Agents transport, load, unload, and assemble items.}
\label{fig:voxel-sim-space}
\vspace{-6mm}
\end{figure}

The voxel-based simulation space abstracts all industrial system components into three types: receptors, agents, and items. Fig. \ref{fig:voxel-sim-space} provides an overview of this architecture. The space is represented as a 3D voxel grid, where each voxel has a unique 3D coordinate. Voxels serve as the basic spatial unit, similar to pixels in 2D space but extended into three dimensions. Each component type is defined as follows:
\begin{itemize}
\item \textbf{\textit{Items}} represent processed materials in the simulated industrial systems. They do not move or change form on their own but can be combined, separated, or nested inside other items as compound items. Each item has a unique ID and, if applicable, tree-structured information describing its composition. 
\item \textbf{\textit{Receptors}} store items and can be stacked. Items inside receptors can be processed by agents. If a receptor is placed above ground level, agents require additional time to load and unload items. Each receptor has a unique ID and may belong to one or more groups, which can define categories such as storage zones.
\item \textbf{\textit{Agents}} are the only active components that move within the simulation space and perform tasks. Their fundamental operations include transportation and assembly. In a single transportation task, an agent loads, transports, and unloads items from one receptor to another. Batch transportation allows an agent to load multiple items from different receptors and unload them at multiple destinations. Assembly operations combine multiple items into a compound item, while disassembly breaks compound items into individual components. It is also possible to process a single item and change its state. It can be represented as an assembly operation that transforms one item into a different item with a new ID. Each agent has a unique ID and an agent-type ID. With different agent types, agents can have different parameters and task-processing mechanisms.
\end{itemize}

As long as the industrial system is engaged in discrete manufacturing or an equivalent process, it can be simulated using the proposed architecture and the RMA-based simulator we prototyped (described later). Discrete manufacturing systems can generally be represented as sequences of processes that combine, separate, or transform materials from initial to final states. In our framework, materials are represented as items, and processes are modeled as interactions between agents and receptors. Application areas include machine tool machining, semiconductor fabrication, solar panel assembly, power transformer production, and rail car manufacturing, to name a few. The following examples illustrate how real-world industrial system components, including but not limited to manufacturing, map to these three categories:
\begin{itemize}
\item {\it Factories}: Items include parts, sub-assemblies, boxes, and pallets. Agents include workers, forklifts, AGVs, and conveyors. Receptors include shelves, storage slots, assembly stations, and floor spaces. 
\item {\it Distribution centers}: Items include merchandise, labels, boxes, and pallets. Agents include workers, AGVs, forklifts, and trucks. Receptors include shelves, packing stations, and floor storage areas.
\item {\it Restaurants}: Items include ingredients, utensils, and dishes. Agents include servers, cooks, and customers. Receptors include tables, kitchen storage, and refrigerators.
\item {\it Ferry terminal operation}: Items include boarding tickets and suitcases. Agents include passengers and cars. Receptors include storage slots on vessels and terminal storage areas.
\end{itemize}

\begin{figure}[h!]
\centering
\parbox{\textwidth}{
\includegraphics[scale=0.61]{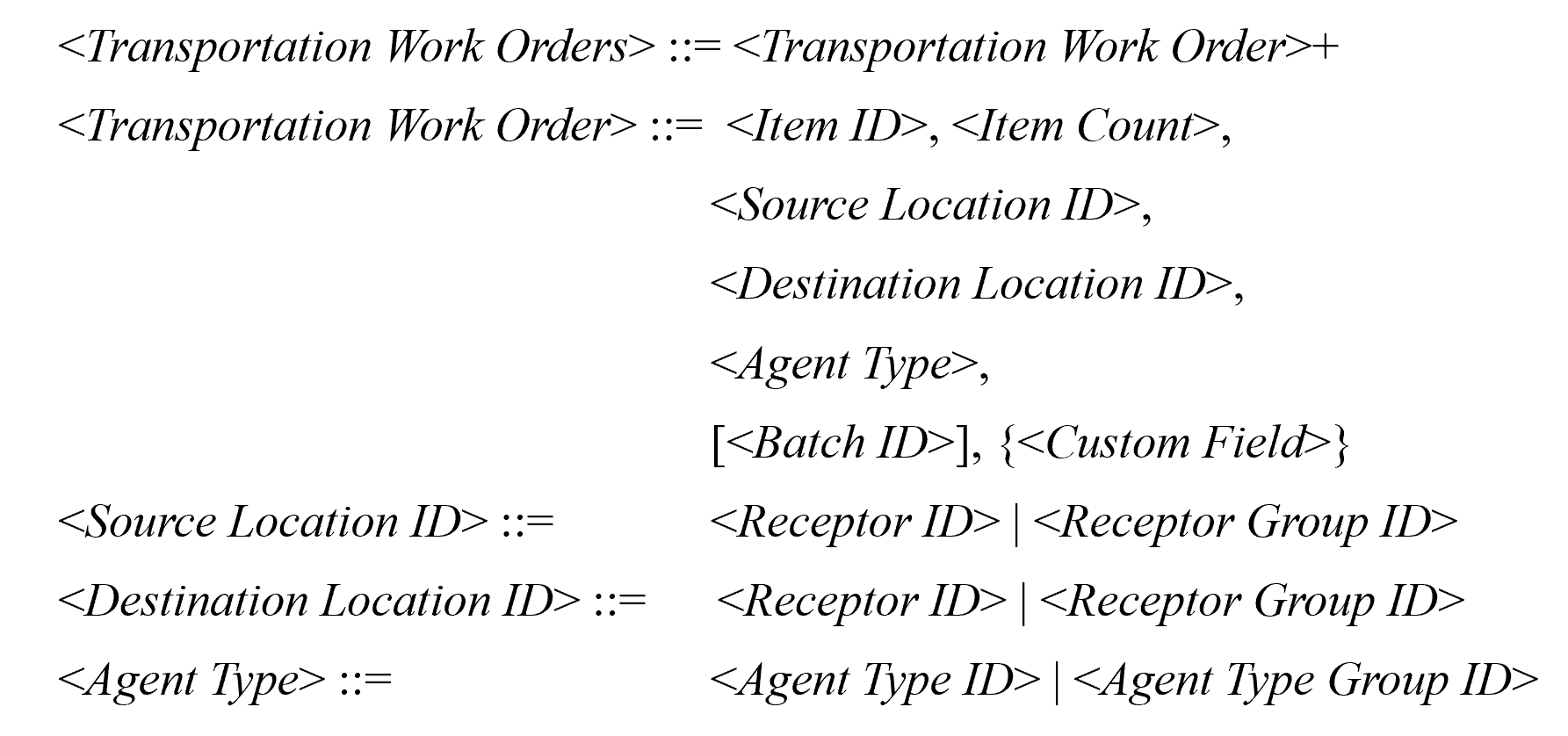}
}
\caption{Data structure definition of transportation work orders.}
\label{fig:t-work-order}
\end{figure}
\begin{figure}[h!]
\centering
\parbox{\textwidth}{
\includegraphics[scale=0.61]{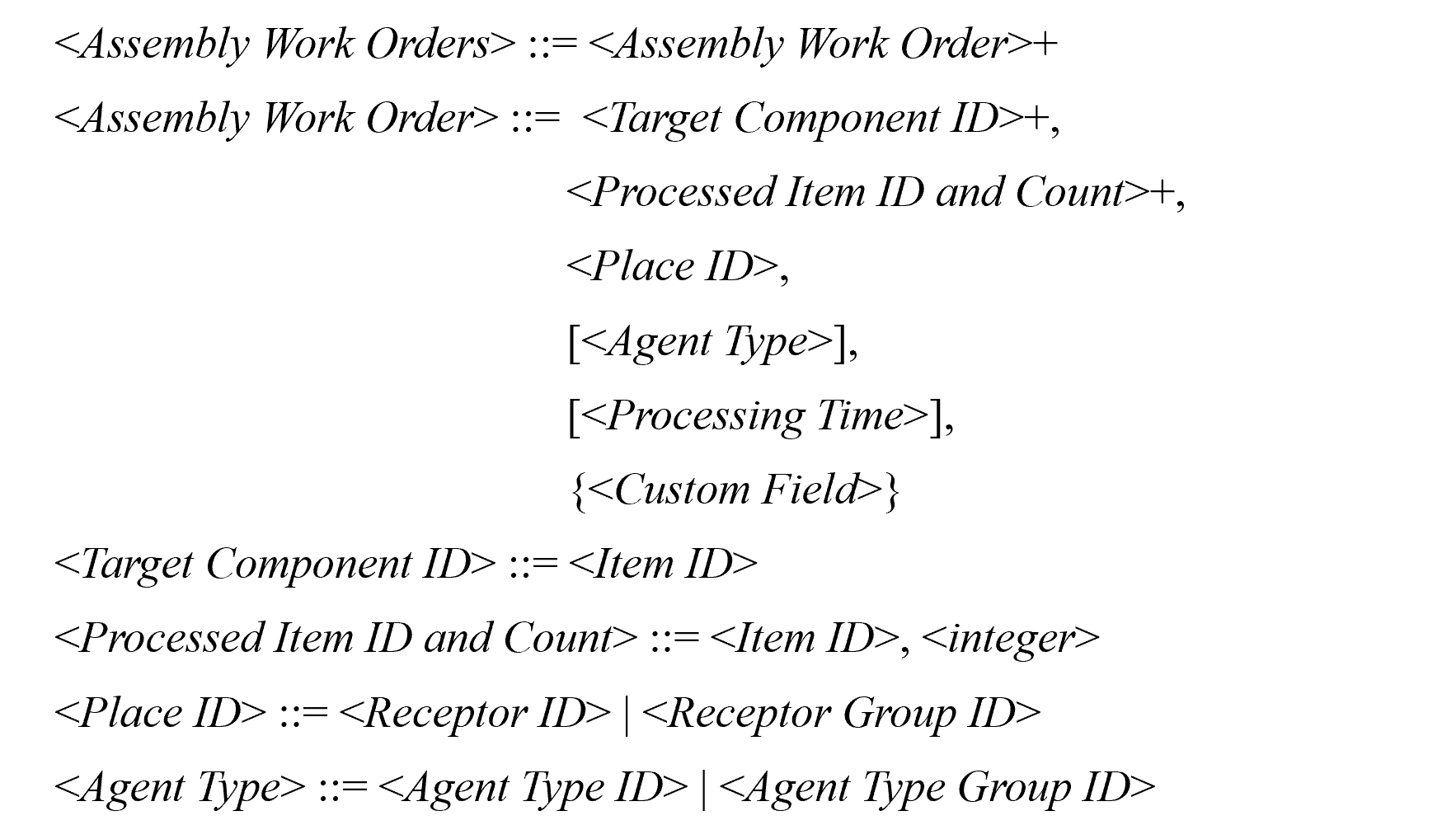}
}
\caption{Data structure definition of assembly work orders.}
\label{fig:assembly-work-order}
\end{figure}
\begin{figure}[h!]
\centering
\parbox{\textwidth}{
\includegraphics[scale=0.61]{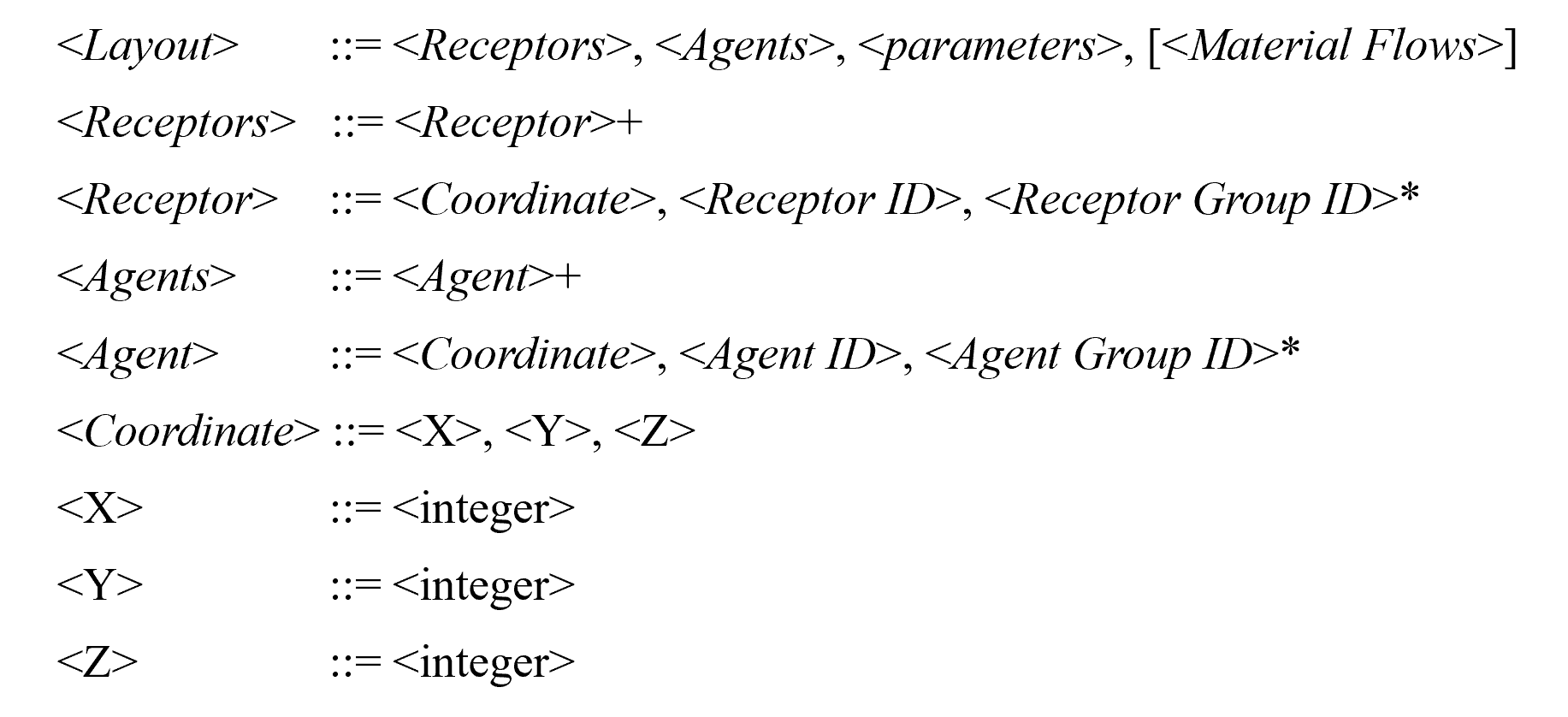}
}
\caption{Data structure definition of layout and basics data.}
\label{fig:layout-basics}
\end{figure}
\begin{figure}[h!]
\centering
\parbox{\textwidth}{
\includegraphics[scale=0.61]{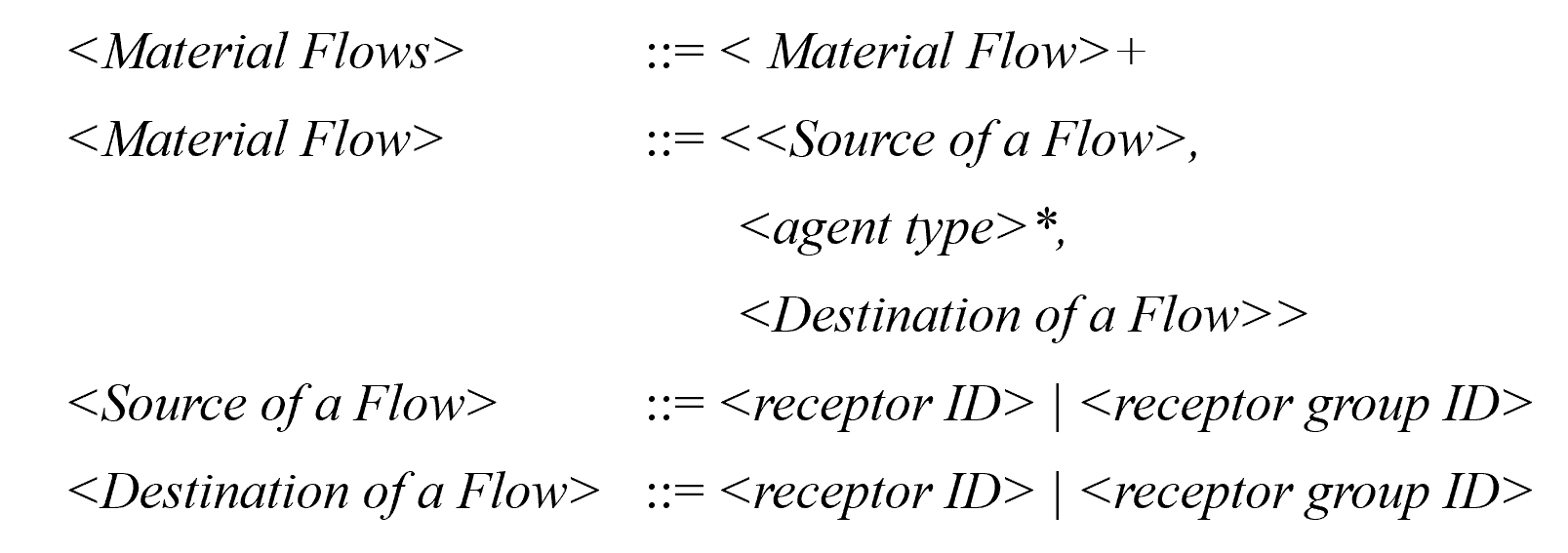}
}
\caption{Data structure definition of material flows.}
\label{fig:material-flows}
\end{figure}
\begin{figure}[h!]
\centering
\parbox{\textwidth}{
\includegraphics[scale=0.61]{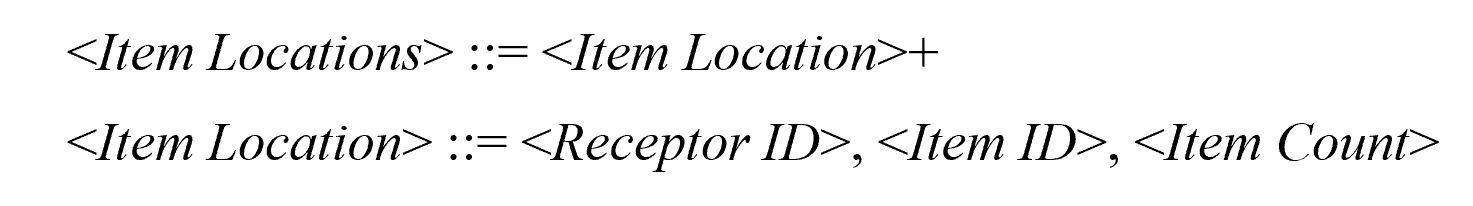}
}
\caption{Data structure definition of item locations.}
\label{fig:item-loc}
\vspace{-4mm}
\end{figure}

\subsection{(S2-1) Distilled Input Data Structure and Semantics}
The input data for the Rapid Modeling Architecture (RMA) consists of three key components: (i) layout and basics data, (ii) work orders for agents, and (iii) item location data. Based on the simulation space architecture presented in the last subsection, these input data define the (i) layout and basics, (ii) tasks for agents, and (iii) item locations at the beginning of simulations. There are two types of work orders: transportation work orders and assembly work orders. The data structure definition of transportation work orders is defined in Fig. \ref{fig:t-work-order}. The data structure of assembly work orders is defined in Fig. \ref{fig:assembly-work-order}. The data structures are defined in EBNF (Extended Backus–Naur Form) styles.

Transportation work orders, as defined in Fig. \ref{fig:t-work-order}, consist of at least one transportation work order. Each transportation work order includes an item ID, item count, source location ID, destination location ID, and the agent type responsible for transportation. The source and destination locations can be either a specific receptor or a group of receptors. Similarly, the agent type can be either a specific type of agent or a group of agent types. Additionally, a transportation work order may include an optional batch ID and more custom fields.

Assembly work orders, as defined in Fig. \ref{fig:assembly-work-order}, consist of at least one assembly work order. Each assembly work order includes one or more target component IDs, one or more processed item ID and count pairs, and a place ID where the assembly occurs. The place can be either a specific receptor or a group of receptors. Optionally, an assembly work order may also specify an agent type, processing time, and one or more custom fields. The agent type can be either a specific type of agent or a group of agent types.

The layout and basics data structure is defined in Fig. \ref{fig:layout-basics}. This data consists of receptors, agents, parameters, and optionally material flows. A valid layout must contain at least one receptor, one agent, and a set of parameters. Receptors represent fixed positions in the system. Each receptor is defined by a 3D coordinate consisting of X, Y, and Z values, a receptor ID, and zero or more receptor group IDs. Similarly, agents represent entities that move or interact within the system. Each agent is defined by a 3D coordinate (initial position), an agent ID, and zero or more agent group IDs. Material flows are defined in Fig. \ref{fig:material-flows}. A material flow represents the movement of items between receptors and is composed of a source, a destination, and at least one agent type responsible for handling the flow. A material flow can originate from either a specific receptor ID or a receptor group ID and must be transported to a designated receptor ID or receptor group ID. 

The data structure of item locations is defined in Fig. \ref{fig:item-loc}. This data consists of at least one item location, where each item location is represented by a receptor ID, an item ID, and the corresponding item count.

\subsection{(S2-2) Flexible Input Data Processing Mechanism based on Self-Order Generation}
While users can define detailed transportation and assembly work orders as explained in the last subsection, these can also be omitted to reduce unnecessary modeling effort. This subsection explains how the simulator maintains flexibility in input data processing through a self-order generation function.

\vspace{1mm}
\subsubsection{Overview of the Extended Discrete-Event Simulator}

\begin{figure}[t]
\centering
\parbox{\textwidth}{
\includegraphics[scale=0.35]{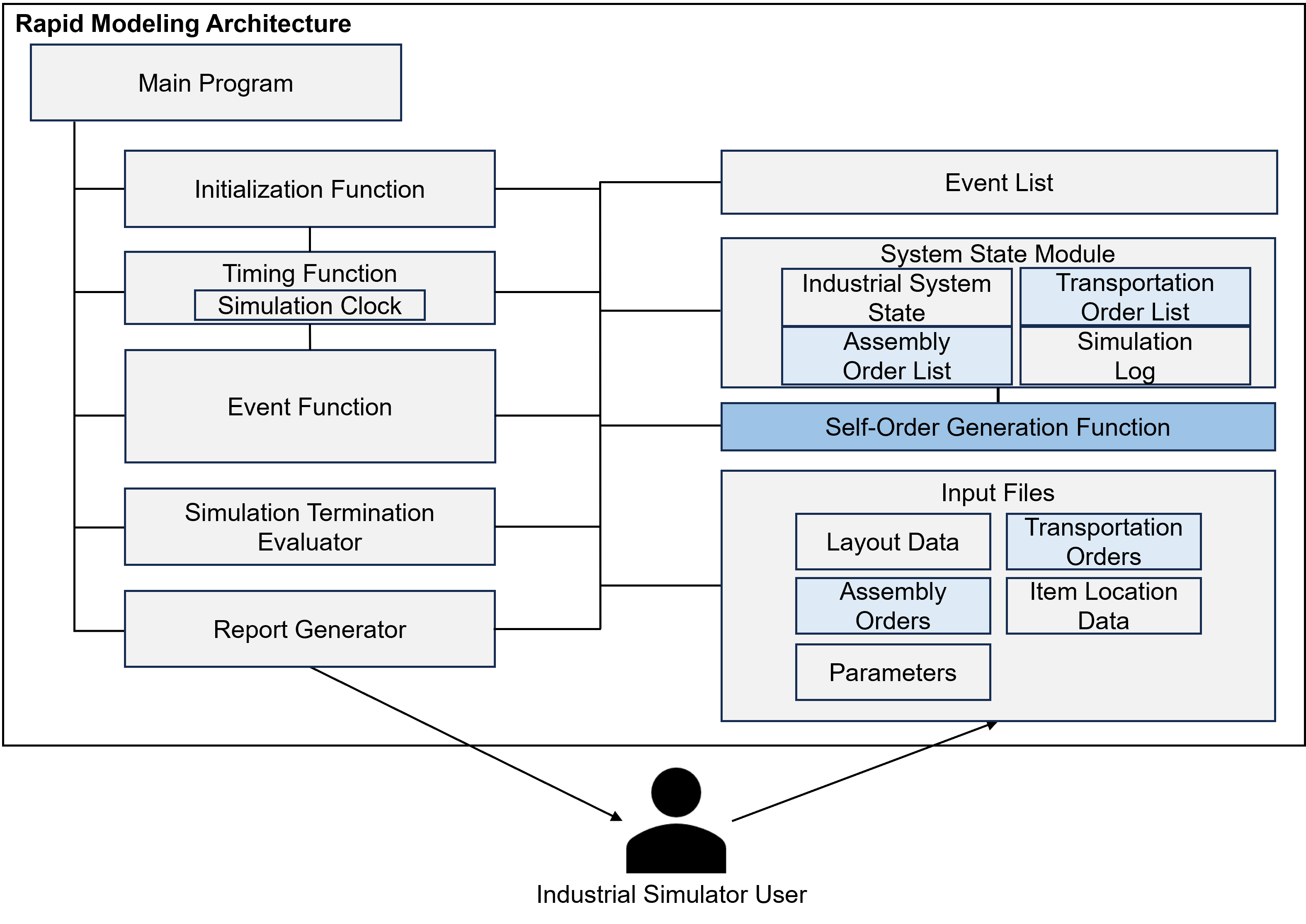}
}
\caption{Extended DES for Rapid Modeling Architecture (RMA).}
\label{fig:rma}
\vspace{-4mm}
\end{figure}

\begin{figure}[t]
\centering
\includegraphics[scale=0.38]{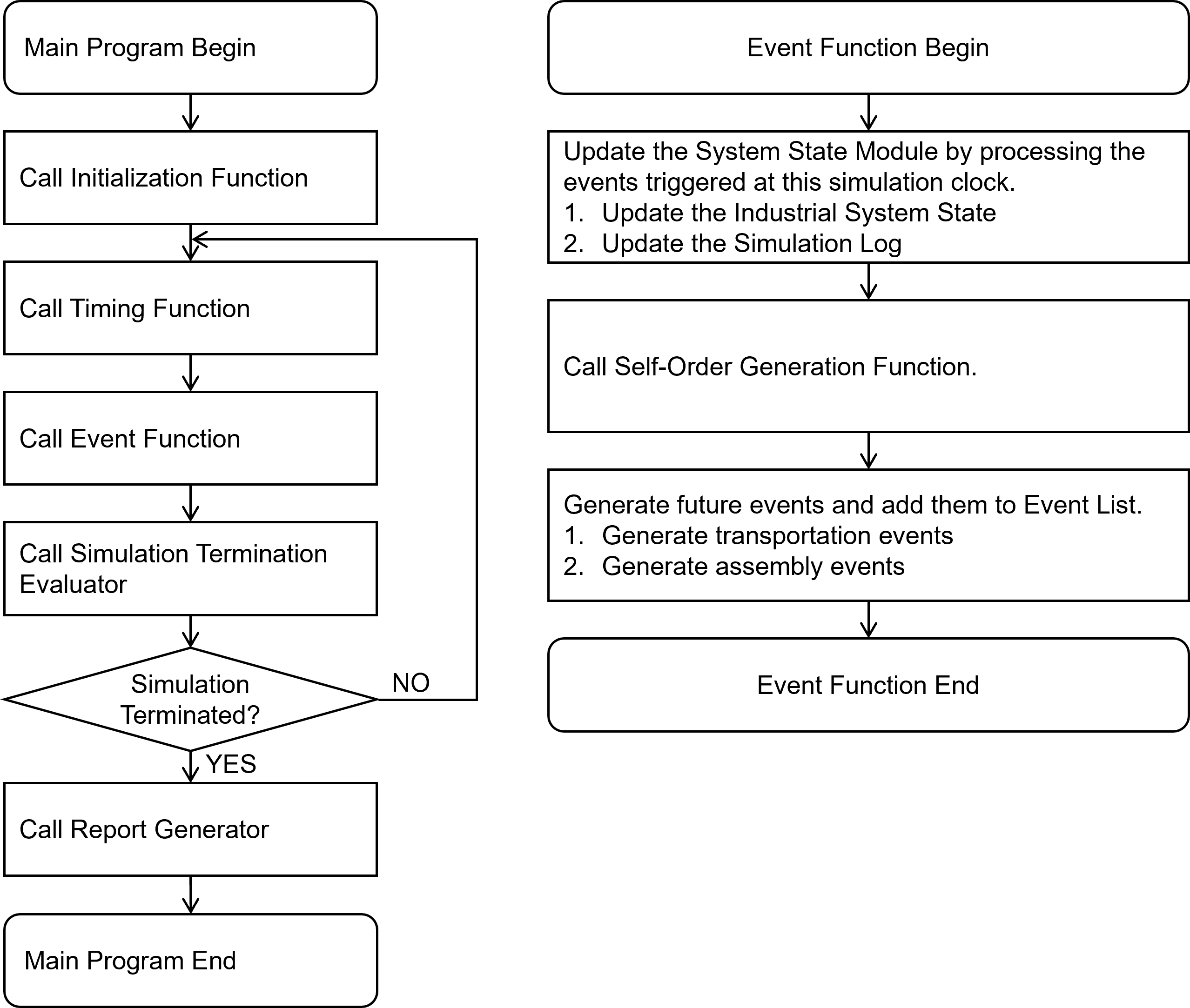}
\caption{Flowchart of the extended DES for RMA.}
\label{fig:flowchart}
\vspace{-4mm}
\end{figure}

The simulator is based on a discrete-event simulation (DES) framework with additional components for dynamic work order generation. The functional architecture is shown in Fig. \ref{fig:rma}, and the simulator’s flow chart is illustrated in Fig. \ref{fig:flowchart}. 

The simulator starts by calling the initialization function that loads the input data, initializes the simulated system state using the data, initializes the event list to be empty, and initializes the simulation clock to be zero in the timing function. The initialization function then creates the events of transportation and assembly based on the system state modules and inserts them into the event list. The timing function is then called to move the simulation clock to the earliest event in the event list. After that, the event function is called. The event function does three things: processing events, calling the self-order generation function, and generating events. First, the event function processes the triggered events, i.e., the transportation events and assembly events that have occurred at the current simulation time, and updates the system state module accordingly. Then, the event function calls the self-order generation function to generate further transportation work orders. Finally, the event function generates events based on the system state and the remaining work orders in the system state module. After that, the simulator calls the simulation termination evaluator to determine if the simulation has reached the terminal state. The terminal state is usually the state with no remaining work orders, but it can be customized. If the simulation is terminated, the simulator will generate a report. If not, the simulator goes back to calling the timing function to advance the simulation clock and proceed with the simulation.

\vspace{1mm}
\subsubsection{Flexibility Provided by the Self-Order Generation Function}

The self-order generation function is a unique extension of the conventional discrete-event simulator (DES). It generates:
\begin{itemize}
\item {\it For Assembly}: Additional transportation work orders that are necessary to make designated components specified in the assembly work orders if parts required to make a component specified in the assembly work orders are not in the receptor or coming to the receptor to perform the assembly. 
\item {\it For Material Flows}: Additional transportation work orders to transport materials based on the material flows specified in the layout and basics data if items are stored in the source receptor of a flow and the transportation work orders for the items are not in the {\it Transportation Order List}.
\end{itemize}
These two simple order generations allow users to avoid preparing the transportation work orders as necessary or preparing the assembly work orders as necessary to simplify the modeling process. Here is how the simulator flexibly adjusts to the input data variations:
\begin{itemize}
\item {\it Material Flows Only}: the self-order generation function generates all the transportation work orders necessary to simulate the transportation.
\item {\it Material Flows and Assembly Work Orders}: the self-order generation function generates all the transportation work orders necessary to simulate the item transportation and assemblies.
\item {\it Material Flows, Assembly Work Orders, and Transportation Work Orders}: the self-order generation function generates the transportation work orders if the transportation for assemblies is not all specified in the transportation work orders in the input data. 
\end{itemize}
Because of the flexibility provided by the self-order generation function, users can omit unnecessary input files to have unnecessary LoDs. 

\section{IMPLEMENTATION OF \\INDUSTRIAL SIMULATOR USING \\PROPOSED RAPID MODELING ARCHITECTURE}
We developed a prototype simulator based on the Rapid Modeling Architecture (RMA) using JavaScript, hosted on a Node.js server. The simulation is visualized using Three.js \cite{threejs2025}.

\subsection{Prototype Architecture}
\begin{figure}[t]
\centering
\parbox{\textwidth}{
\includegraphics[scale=0.32]{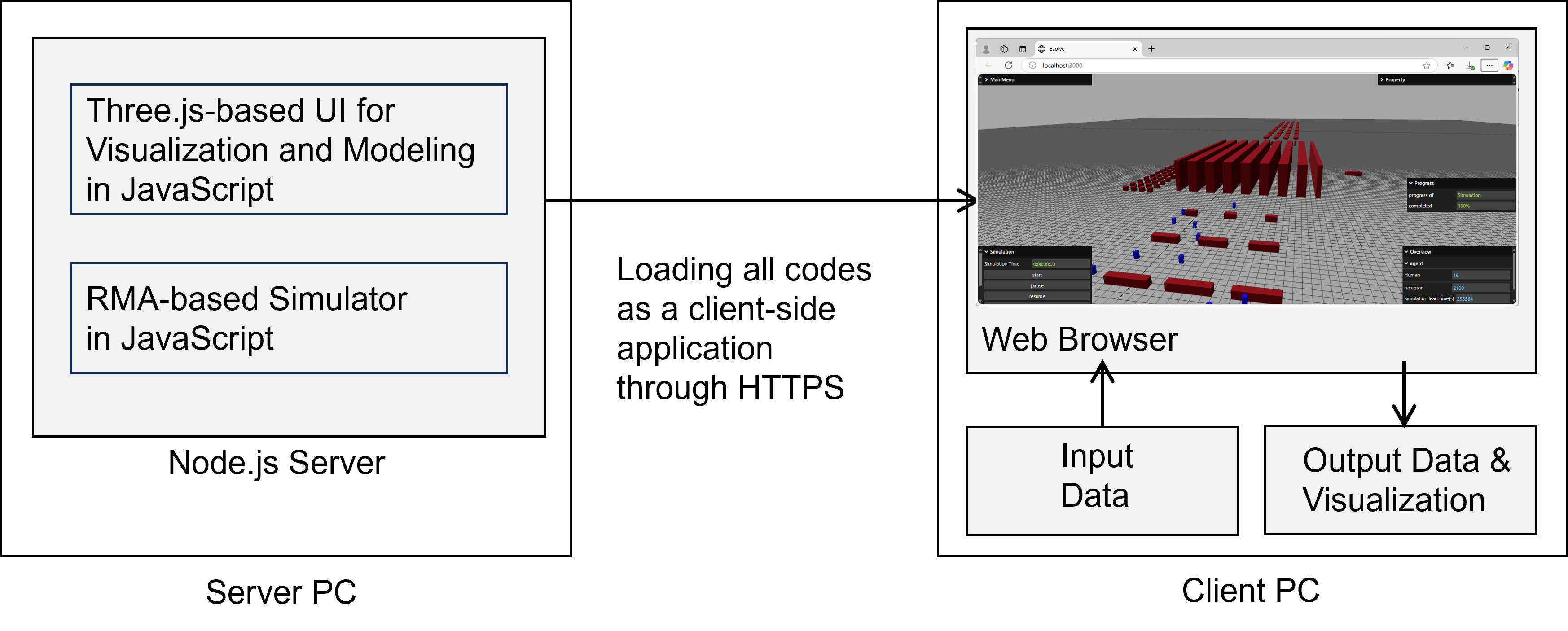}
}
\caption{Implemented architecture of RMA-based simulator.}
\label{fig:impl-arch}
\end{figure}

The architecture of the prototype is shown in Fig. \ref{fig:impl-arch}. It follows the RMA-based structure depicted in Fig. \ref{fig:rma}, and the internal data structure of the simulator adheres to this architecture. The Three.js-based user interface handles both visualization and modeling. It receives simulation results from the simulator module and displays them through the interface. Both modules are implemented as client-side applications. When a user accesses the application hosted on the Node.js server, the browser loads both the visualization and modeling interface along with the RMA-based simulator. This allows the user to run and view the simulation locally in the browser.

\subsection{Data Structure}
\begin{figure}[t]
\centering
\includegraphics[scale=0.30]{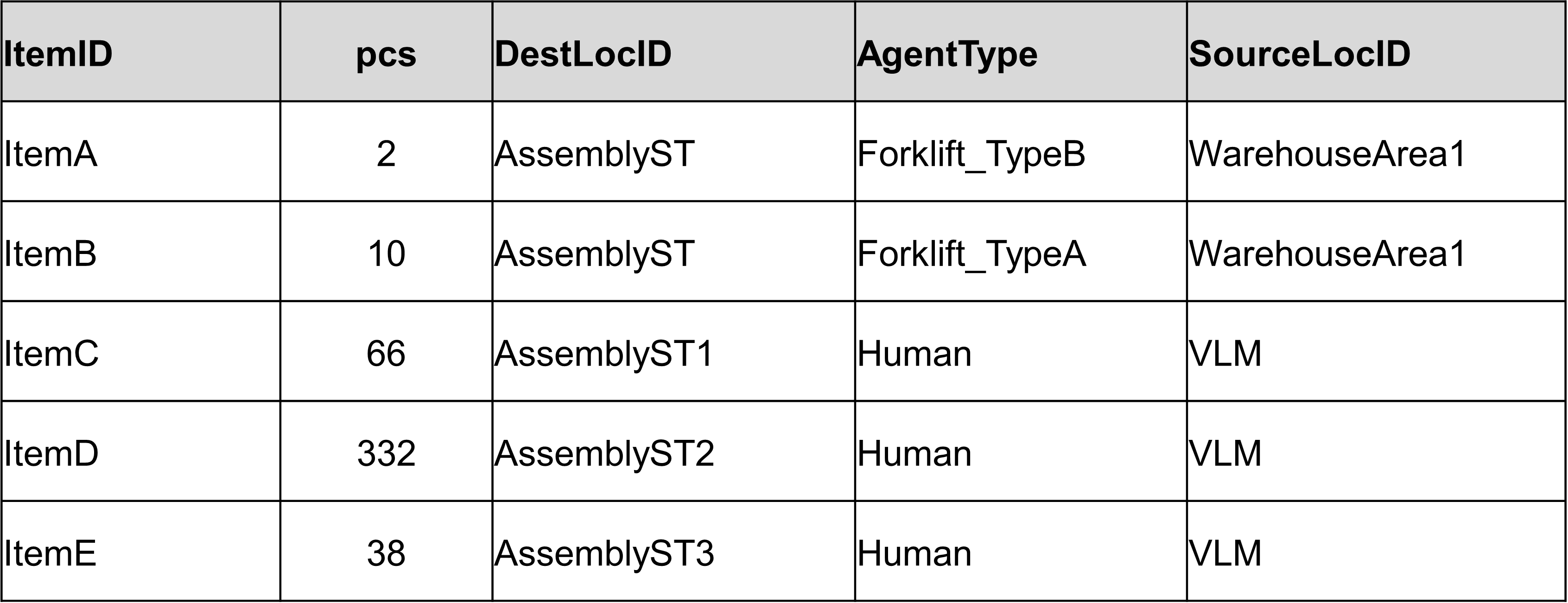}
\caption{Structure and example of implemented transportation work orders.}
\label{fig:example-t-orders}
\vspace{-4mm}
\end{figure}

Fig. \ref{fig:example-t-orders} presents the structure and examples of implemented transportation work orders in CSV (Comma-Separated Values). The columns are ItemIDs, pcs, destination location ID (DestLocID), source location ID (SourceLocID), and agent types (AgentType). The source location IDs and destination location IDs can be a receptor ID or a group ID of receptors. Each row in Fig. \ref{fig:example-t-orders} represents a transportation work order for an agent to perform. The first line represents a task for an agent with ‘Forklift\_TypeB’ agent type to transport two pieces of ‘ItemA’s from ‘WarehouseArea1’ to ‘AssemblyST’. In this case, ‘WarehouseArea1’ is a group ID of receptors in the warehouse area one. ‘AssemblyST’ is also a group ID of receptors representing the stations in the assembly area. The selection of receptors in each group can be customized based on the analysis and operation requirements. If users want to process the tasks in a batch, batch ID column can be added to the table data in Fig. \ref{fig:example-t-orders} to group the tasks. 

\begin{figure}[t]
\centering
\includegraphics[width=0.6\linewidth]{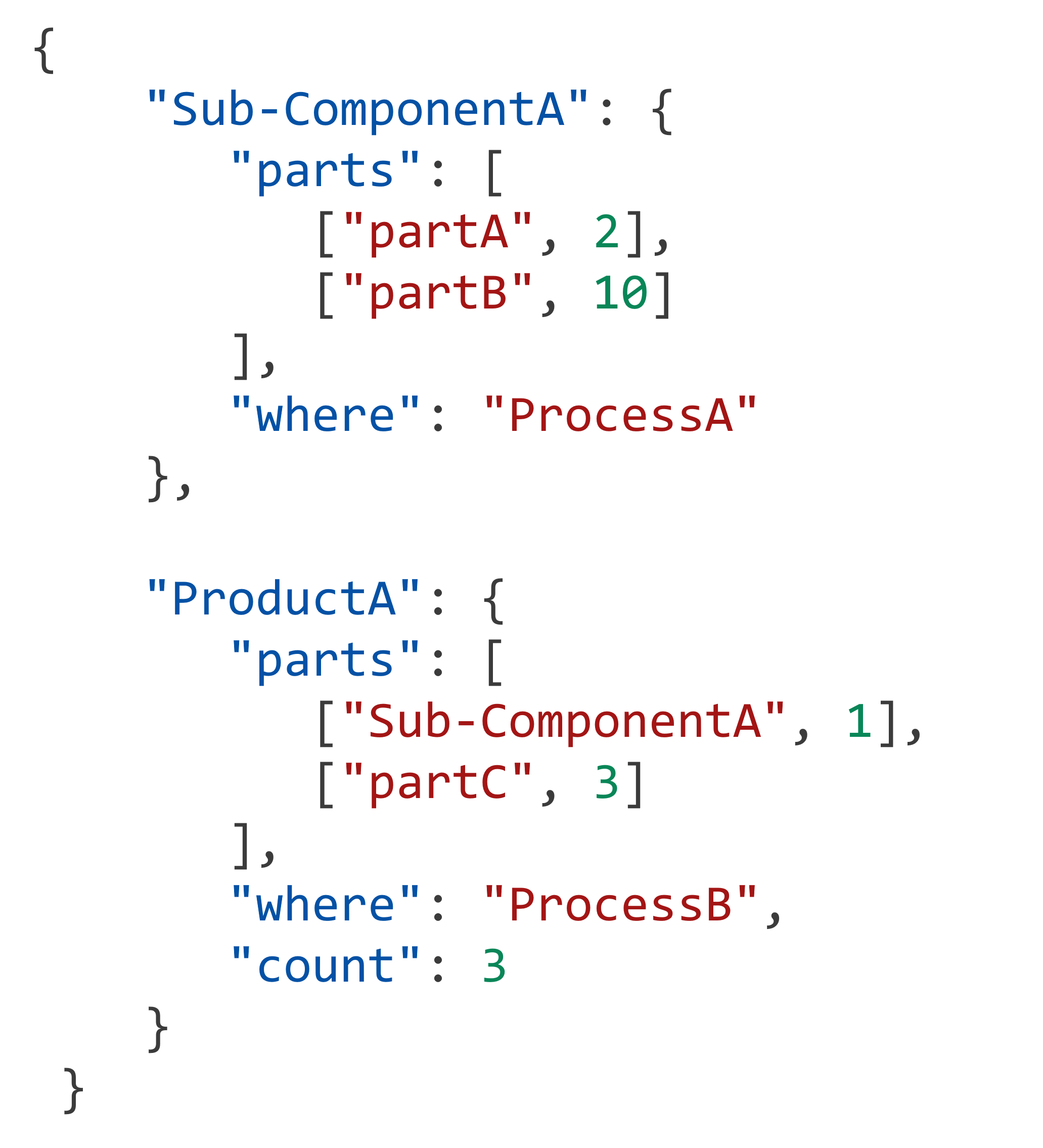}
\caption{Structure and example of implemented assembly work orders.}
\label{fig:example-assembly-orders}
\vspace{-4mm}
\end{figure}

Assembly work orders are implemented as tree-structured data that represent composition of outcomes, e.g., products, dishes, etc. Fig. \ref{fig:example-assembly-orders} shows the implemented structure and examples of assembly work orders in JSON (JavaScript Object Notation).
"ProductA" represents the ID of a final product. It has nested key-value pairs that define its "parts", assembly location ("where"), and production quantity ("count"). Under "parts", "ProductA" consists of one unit of "Sub-ComponentA" and three units of "partC". The "where" key specifies "ProcessB", meaning the final assembly occurs in ProcessB. The "count" key is 3, indicating that three units of "ProductA" are expected to be manufactured. Since each "ProductA" requires one "Sub-ComponentA", this also means that three units of "Sub-ComponentA" need to be produced to meet demand.
As long as the semantics are retained, the transportation work orders can be in other file formats, such as JSON files, Extensible Markup Language (XML) files, etc.

%
%

\vspace{-2mm}
\subsection{User Workflow for Simulation Modeling in the RMA-based Simulator}
The user’s workflow can be defined as follows:
\begin{enumerate}
\item {\it Gathering requirements}: Define the simulation goal, such as evaluating throughput or layout efficiency. Identify system components (items, agents, receptors), available data, and necessary modeling granularity.
\item {\it Placing Objects}: Create a voxel-based layout by placing receptors and agents using the Three.js-based interface. Assign group IDs to structure areas (e.g., zones, processes). Set parameters to define the basic characteristics.
\item {\it Behavior Modeling}: define material flows using the interface, generate the input data of transportation and assembly work orders. There three dominates the behavior modeling, but if there is a need to customize the behavior, the user can modify the Event Function and/or Self-Order Generation Function in RMA to tailor the behavior (e.g., handling exceptions, specific rework flow definition outside of regular operations, simulating exceptions like facility downtime, etc.)
\item {\it Run simulations, analyze, and improve model}: Load the model in the browser to run and visualize the simulation. Use the results to identify issues, test alternatives, and refine input data. The quick iteration cycle enables rapid feedback for early-phase decision-making. 
\end{enumerate}

\section{EVALUATION OF PROPOSED RAPID MODELING ARCHITECTURE IN INDUSTRIAL SYSTEM OPTIMIZATION}
This research aims to reduce the time required to model industrial systems. To assess the effectiveness of our RMA-based simulator, we conducted a modeling experiment comparing it with a widely used proprietary simulator to measure the reduction in modeling time. The following subsections describe the experimental setup and results.

\subsection{Target Industrial System for Modeling and Analysis}
\begin{figure}[t]
\centering
\includegraphics[scale=0.38]{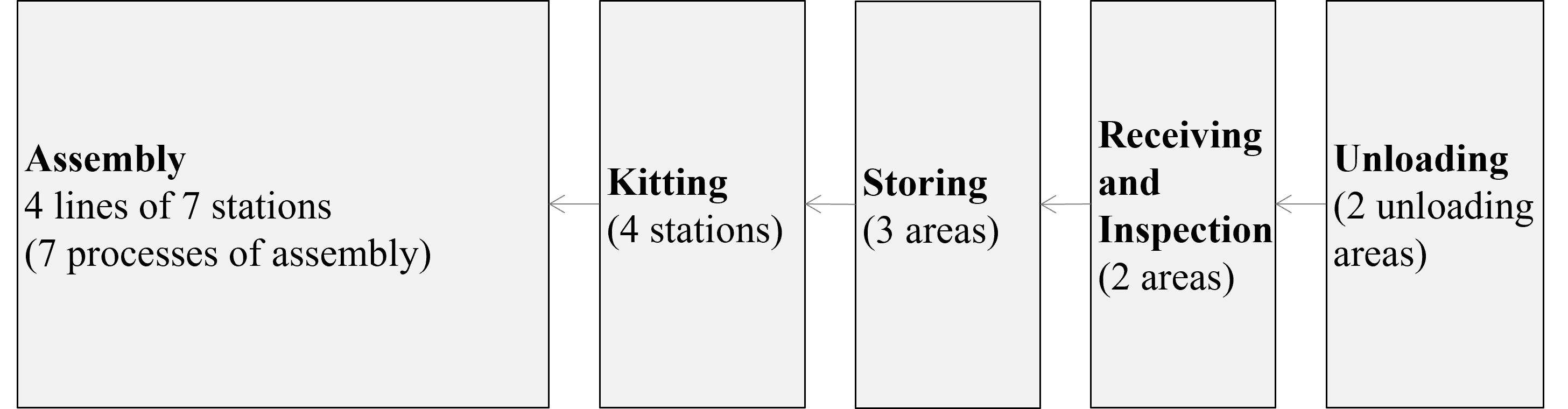}
\caption{Line configuration of modeled rail car factory. There are five processes, and assembly process has seven stations per line. There are four lines of stations in the assembly process. Unloading starts from the trucks arriving at the factory. This line is for the manufacturing one type of rail car.}
\label{fig:factory-process}
\vspace{-5mm}
\end{figure}

As a target of the modeling experiment, we chose a rolling stock factory (rail car factory) based on an actual plant since it is one of the typical industrial systems. The processes we focus on start with unloading from trucks and end with final inspections of rail cars. The model's objective is to evaluate the makespan of rail cars, i.e., how long it takes to finish manufacturing a rail car from the beginning. The components need to include human workers, multiple types of forklifts, pallets, rail car parts, tools, shelves, assembly stations, AGVs, etc. The process configuration of the target rail car factory is shown in Fig. \ref{fig:factory-process}. There are four main processes, namely, unloading, receiving and inspection, storing, kitting, and assembly. There are two unloading areas in the unloading process, two areas in the receiving and inspection process, three types of areas in the storing process, four stations in the kitting process, and finally, four lines in the assembly process where each line has seven stations. 

This modeling aims to analyze the system performance [cars/month] and make decisions on machinery procurement as to how many forklifts and AGVs are needed to achieve the system performance goal.

\subsection{Evaluation Method}
We have chosen Visual Components (VC) 4.10 to compare since it is commonly used for modeling in the early phase of the projects. We first designed the rail car factory layout and operation. Then, we modeled the rail car factory using the prototyped RMA-based simulator and Visual Components while measuring the modeling time of every step. We assigned this experiment to a single developer with experience using Visual Components and RMA-based simulators. We set the limit of modeling time to 20 hours, as it is impractical to have more time in the early decision-making process phases.

\subsection{Measured Modeling Time}
\begin{figure}[t]
\centering
\parbox{\textwidth}{
\includegraphics[scale=0.30]{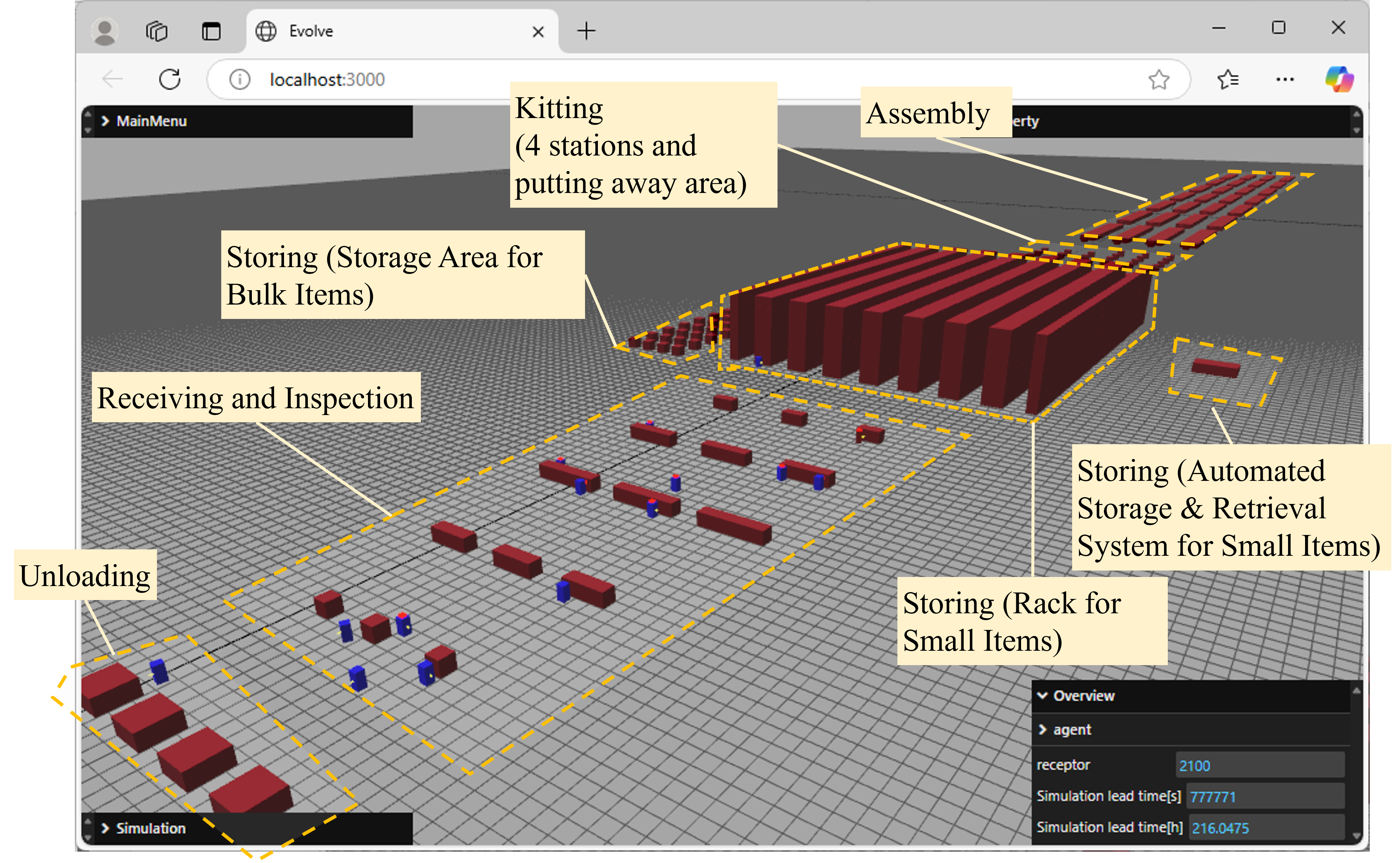}
}
\caption{Oveview of modeled rail car factory based on the prototyped RMA-based simulator. There are receptors representing components in each area, and agents carrying things among them. The small red boxes on the agents represent items being carried by the agents.}
\label{fig:model-overview.png}
\vspace{-1mm}
\end{figure}

The overview of the modeled rail car factory using the prototyped RMA-based simulator is shown in Fig. \ref{fig:model-overview.png}. In each process of the rail car factory, receptors and agents are placed based on the target system's layout. In the unloading area, each receptor represents a truck coming into the factory for unloading. 
The item location data has the list of items that these trucks are expected to contain at the beginning of the simulation.
In the receiving and inspection process, receptors represent tables and floor spaces where items are stored and inspected. In the storing process, there are three types of storage: floor space for bulk items, racks for small items, and an automated storage and retrieval system for small items. After storing, there is a kitting process with four stations and floor space for putting away. Finally, four assembly lines and stations complete the assembly and finalization of the products. AGVs partially perform the transportation from the kitting and to assembly stations.

As a result, the RMA-based simulator shortened the modeling time by 78.3\%. Table \ref{table:result-table} shows the modeling time measurements. The reduction is significant because it used to take a week to model the target industrial system, but with an RMA-based simulator, it only takes half a day. In a month or so, people need to make several decisions based on the simulation results. Given this time reduction in modeling, almost all decisions can be supported, which accelerates and improves decision-making. The modeling time consists of two categories: placing objects and modeling behavior. Modeling behavior has two categories: setup and customize. During the object placement, visual components took more time as we needed to specify and set up objects to be used in the simulation. Modeling behavior also took more time for VC to set up and customize the flow of materials using various types of objects.
\begin{table}[tb]
\caption{Modeling Time Comparison Between Simulators}
\label{table:result-table}
\vspace{-2.5mm}
\begin{center}
\includegraphics[scale=0.60]{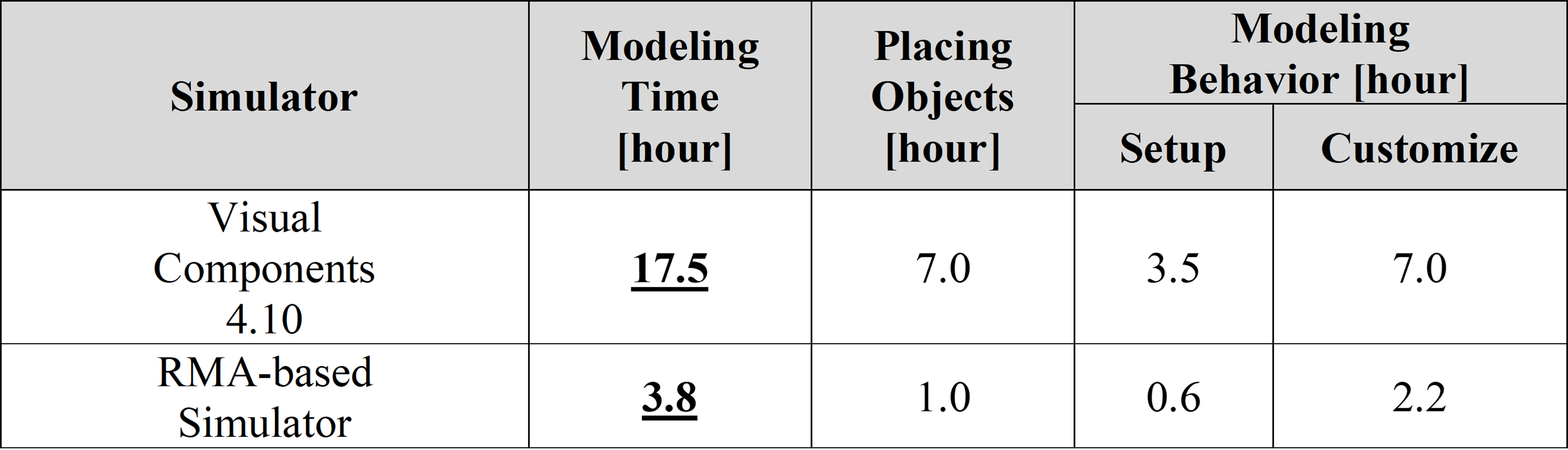}
\end{center}
\vspace{-6mm}
\end{table}

Because of the nature of the RMA-based simulator, the system's behavior is more fine-grained, as we incorporated the inventory master data to generate item location data and M-BOM (Manufacturing BOM) to generate assembly data. Incorporation of these data can be done in Visual Components as well; however, given the time constraints, it was impractical to custom code the incorporation capabilities. If we stopped the modeling using the RMA-based simulator with the same accuracy of behavior as Visual Components, the time for the RMA-based simulator's modeling behavior would be about half.

\subsection{Issues Found and Solved Through Simulations}
\begin{itemize}
\item \textbf{\textit{Collision risks}}: while we were showing the simulations on the Three.js-based visualization interface, we noticed that certain parts of the system get crowded quite often. We then customized the RMA-based simulator to count the collision risks and created a heat map of the risks. We took this information into account when designing the layout and improved safety.
\item \textbf{\textit{Rack size reduction}}: while we were simulating the operations with different sizes of rail car manufacturing orders, we noticed that the occupancy of racks in the warehouse was less than expected. Even though it is important to have extra space in the inventory, we have decided to reduce the number of racks as it was important to improve safety by allocating more space to a certain area to reduce the collision risks.
\end{itemize}
Through these findings on the side, as well as the original system performance evaluation of makespan while changing the number of workers, AGVs, and forklifts, we have confirmed that our RMA-based simulator preserved the essential details required to find meaningful issues to improve the designs in the early phases.

\section{CONCLUSION}
To accelerate decision-making and improve industrial system design in its early phases, we proposed a Rapid Modeling Architecture (RMA) that reduces modeling time while preserving the essential details for meaningful analysis. Our contributions include:
\begin{itemize}
\item Proposing the RMA to shorten the modeling time while maintaining the essential details for identifying key issues in industrial systems.
\item Developing a prototype simulator based on the RMA. 
\item Modeling an actual factory layout and operation using the prototype to analyze and optimize the system.
\item Comparing modeling time between the RMA-based simulator and a conventional simulator to quantify the modeling time reduction.
\end{itemize}
Our future work includes applying RMA on top of another simulator to evaluate how RMA can simplify the modeling, using an RMA-based simulator for the RL studies as the modeling cost often stands in the way of optimizing industrial systems with RL.
Another possible workstream is to connect the RMA-based simulator to the digital thread of the digital twin-based system in order to forecast bottlenecks and risks.

\addtolength{\textheight}{-12cm}   








\bibliographystyle{IEEEtran}
\bibliography{references}


\end{document}